\documentclass[twocolumn, tighten, trackchanges, resetfootnote]{aastex7}

\usepackage{amsmath}
\usepackage{rotating}

\shorttitle{Proper 3D Modeling of Wide Binary Orbits}
\shortauthors{Chae}

\begin{document}

\title{On the Proper 3D Modeling of Orbits in Wide Binary Gravity Tests}

\author[orcid=0000-0002-6016-2736, gname=Kyu-Hyun, sname=Chae]{Kyu-Hyun Chae}
\email{chae@sejong.ac.kr}\thanks{corresponding author: chae@sejong.ac.kr \\ kyuhyunchae@gmail.com}
\affiliation{Department of Physics and Astronomy, Sejong University, 209 Neungdong-ro Gwangjin-gu, Seoul 05006, Republic of Korea}
\author[gname=Hyoseok, sname=Park]{Hyoseok Park}
\affiliation{Department of Physics, Chungnam National University, 99 Daehak-ro, Yuseong-gu, Daejeon 34134, Republic of Korea}
\email{phs137@o.cnu.ac.kr}
\author[orcid=0000-0002-0096-4702, gname=Youngsub, sname=Yoon]{Youngsub Yoon}
\affiliation{Department of Physics and Astronomy, Sejong University, 209 Neungdong-ro Gwangjin-gu, Seoul 05006, Republic of Korea}
\email{youngsuby@gmail.com}

\begin{abstract}
A gravitationally-bound system of wide binary stars with precise 3D relative displacement $\mathbf{r}$ and velocity $\mathbf{v}$ can allow a Keplerian orbit reconstruction for an assumed gravitational parameter $G$. A sufficiently large sample of such systems can also be used to directly infer $G$ specific to the sample. A recent study argued that treating the semi-major axis ($a$) explicitly free with an empirical prior imposed on it can dramatically change the inference of $G$ in the Chae et al.\ sample of 36 wide binaries. Here we introduce and test a Bayesian Markov chain Monte Carlo algorithm that allows $a$ to be explicitly free (with or without a prior on $a$), and find using the $\alpha$-Cen AB - proxima system and the Chae et al.\ sample that the explicit-$a$ algorithm with an observational constraint is fully equivalent to the implicit-$a$ algorithm with an internal consistency relation involving the sky-projected separation $s$. For the $\alpha$-Cen AB - proxima system, we present orbits in Milgromian (MOND) as well as Newtonian gravity models. We show that a recent claim of Newtonian gravity from the Chae et al.\ sample is a consequence of a fatal error in the code or arbitrarily enlarged errors. We discuss how wide binaries can be properly analyzed and modeled to probe gravity at low acceleration and test gravity models, including the necessity of deriving individual-specific parameters, the importance of follow-up observations for anomalous systems, and the relevance of Bayesian MCMC approaches.
\end{abstract}


\section{Introduction}\label{sec:intro}

An observed wide binary system of stars is believed to be at a phase of its gravitationally-bound orbit. Although measurements of the 3D relative displacement $\mathbf{r}$ and velocity $\mathbf{v}$ between the two stars can represent only an instantaneous state, its underlying orbit can, in principle, be reconstructed with a sufficiently precise set of ($\mathbf{r}$, $\mathbf{v}$) assuming that the orbit is Keplerian and gravitational parameter $G$ takes a certain value such as the Newtonian value $G_{\rm N}$. This is because the number of the set ($\mathbf{r}$, $\mathbf{v}$) equals the number of orbit free parameters. A well-known example is provided by Proxima's orbit around $\alpha$-Cen AB (barycenter) as constructed by \cite{Kervella:2017} and \cite{Akeson:2021} assuming Newtonian gravity.

While wide binaries can be used to infer Newtonian orbits when the observed value of $v(\equiv|\mathbf{v}|)$ is less than the Newtonian escape velocity $v_{\rm escN}$ as in the $\alpha$-Cen AB-Proxima system, wide binaries may also be used to infer $\gamma (\equiv G/G_{\rm N})$ as recently introduced by a couple of studies \citep{Chae:2025,Chae:2026}. These studies use the logarithmic parameter $\Gamma (\equiv \log_{10}\sqrt{\gamma})$, which will also be used throughout in this work. Algorithms to infer $\Gamma$ are based on the realization that there is a degeneracy between $\Gamma$ and a set of orbit parameters (i.e., eccentricity $e$, true anomaly $\Delta\phi$, and semi-major axis $a$ or 3D separation $r$) through the vis-viva equation, and the observed set ($\mathbf{r}$, $\mathbf{v}$) constrains the orbit parameters to some extent. Because the constraining power of ($\mathbf{r}$, $\mathbf{v}$) is not strong, the inferred probability density function (PDF) of $\Gamma$ for an individual wide binary is usually so broad that standard and nonstandard gravities cannot be discriminated.  

Nevertheless, the \cite{Chae:2026} algorithm to infer $\Gamma$ is useful in two important ways. First, the algorithm can be used to infer boosted gravity in an apparently anomalous pair, which appears to form an isolated binary system gravitationally associated with each other in the 3D space based on its observed characteristics (see \citealt{Chaeetal:2026}) but cannot be gravitationally bound with $v>v_{\rm escN}$ in Newtonian gravity. If $\Gamma=0$ is imposed, such an anomalous system does not allow a viable orbit and thus cannot be analyzed in any meaningful way. Second, when a sufficiently large sample of $N$ wide binaries is available, individual PDFs of $\Gamma$ can be statistically consolidated to derive a much narrower PDF with the width proportional to $1/\sqrt{N}$.

The \cite{Chae:2026} algorithm is a generalization of the standard parameter optimization procedure of fitting six model parameters of $\mathbf{\Theta}_{\rm Newton}=\{a,e,i,\phi_0,\Delta\phi,\Omega \}$ (in Newtonian gravity) to the observed set ($\mathbf{r}$, $\mathbf{v}$). Here, $i$ is the angle of inclination, $\phi_0$ is the argument of the periastron, and $\Omega$ is the argument of the ascending node, while the other symbols have been defined above. Because the fitting procedure uses the same six mathematical relations (whose parametric forms may vary depending on the author) between the parameters and the measured quantities, the same data should always yield the same values of $a$, $e$, and $\Delta\phi$, but the values of the other angles depend on the frame where $\mathbf{r}$ and $\mathbf{v}$ are expressed. 

When the equatorial coordinate system is used to express $\mathbf{r}$ and $\mathbf{v}$ (since it is a Cartesian frame to a good approximation for the 3D space containing the binary system once the spherical geometric (perspective) effects are corrected for), the two components $r_{\rm R.A.}$ and $r_{\rm decl.}$ of $\mathbf{r}$ are extremely precisely measured by Gaia and thus they can be used as direct inputs into the mathematical (internal consistency) relations reducing the number of free parameters by two. In particular, the ratio $r_{\rm decl.}/r_{\rm R.A.}$ is essentially exact because the distance factor cancels out while the sky-projected physical separation $s(\equiv\sqrt{r_{\rm R.A.}^2+r_{\rm decl.}^2} )$ of the second star (component B) from the first star (component A which is at the origin of the frame) inherits the uncertainty of $d_A$ (distance of A) which can be sufficiently small (e.g., $\la 0.1\%$ for the \cite{Chaeetal:2026} sample). The \cite{Chae:2026} algorithm directly uses $r_{\rm decl.}/r_{\rm R.A.}$ and $s$ as fixed inputs to the mathematical relations, which determine the semi-major axis $a$ and the argument of the ascending node $\Omega$ from the other parameters.

\cite{SaadTing:2026} have recently questioned the \cite{Chae:2026} algorithm for using a mathematical relation of parameters including a fixed $s$ to reduce the number of free parameters. They argue that treating $a$ as a free parameter along with an empirical prior rather than letting it be determined by the mathematical relation can dramatically change gravity inference. Their claim is unexpected because the mathematics of Keplerian orbits is fixed and the tiny measurement error of $s$ ensures that it can be used as a fixed input into the mathematics, as pointed out above. 

Considering the unexpected claim by \cite{SaadTing:2026}, here we consider a purposely extended algorithm in which the semi-major axis is explicitly treated as a free parameter and the measured sky-projected separation $s$ (along with its measurement uncertainty) is included in a likelihood (but not used as a constant in a mathematical relation), and present Bayesian results for example binary systems including the $\alpha$-Cen - Proxima system and systems from the \cite{Chaeetal:2026}. 

In this work, we carefully address the question of what the proper modeling of wide binary orbits should be in the posed problem of inferring gravity at low acceleration and testing gravity models using wide binaries. It will be explicitly shown that the introduction of a free parameter for the semi-major axis has negligible effects in the Bayesian inference of orbit parameters and $\Gamma$ compared with the implicit-$a$ algorithm of \cite{Chae:2026}. We will then show that the \cite{SaadTing:2026} claim of a Newtonian result from the \cite{Chaeetal:2026} sample is due, in fact, to a fatal error in the code, or a consequence of arbitrarily enlarged errors for the measured quantities, in particular the sky-projected separation between the two stars. In this work, we will also explicitly show that the parameters of Proxima's orbit inferred with the \cite{Chae:2026} algorithm fully agrees with the \cite{Kervella:2017} inference although two studies use completely independent algorithms and work in different reference frames. Then, we will present orbits obtained for MOND \citep{Milgrom:1983,Milgrom:2010} gravities as previous studies considered only Newtonian gravity.

The outline of the paper is as follows. We start with a brief description of the extended algorithm that includes the semi-major axis as a free parameter, highlighting only the difference from the \cite{Chae:2026} algorithm (Section~\ref{sec:method}). In Section~\ref{sec:result}, we present results with the extended algorithm and compare them with other results including those with the \cite{Chae:2026} algorithm and Proxima's orbit inferred by \cite{Kervella:2017}. We discuss and conclude in Section~\ref{sec:conc}. In Appendix~\ref{sec:coordtranf}, we discuss how the perspective effects of the spherical geometry can be accurately accounted for through coordinate transformations. The extended algorithm treating the semi-major axis as a free parameter is available on Zenodo under an open-source Creative Commons Attribution license \dataset[]{https://doi.org/10.5281/zenodo.23857950}.

\section{An Extended Algorithm with a Free Semi-major Axis} \label{sec:method}

\begin{figure}[!htb]
    \centering
    \includegraphics[width=1.05\linewidth]{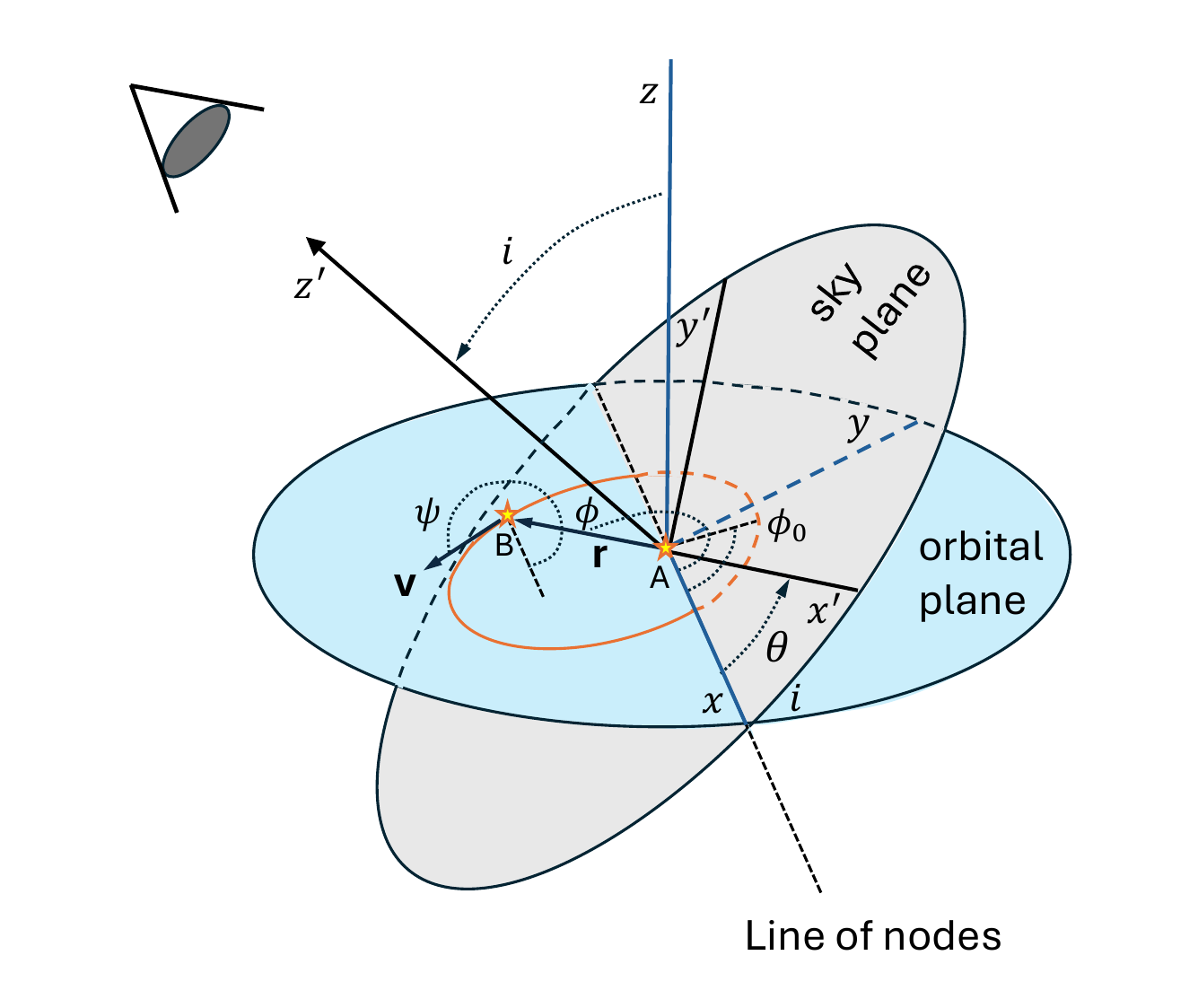}
    \caption{(Adapted from \cite{Chae:2026}) A 3D geometry of an elliptical orbit of a binary expressed in the equatorial frame. The observer's frame is described by the $x^\prime y^\prime z^\prime$ coordinate system, where $x^\prime$ is in the negative R.A. direction, $y^\prime$ is in the positive decl.\ direction, and $z^\prime$ is in the direction towards the observer.
    }
    \label{fig:Euler2angles}
\end{figure}

Figure~\ref{fig:Euler2angles} shows a general 3D configuration of an elliptical orbit of a binary system in the equatorial frame. Under the assumption that the system is gravitationally bound and the orbit is Newtonian or pseudo-Newtonian with a general gravitational parameter $G$, the six components of the relative displacement $\mathbf{r}=(x^\prime,y^\prime,z^\prime)$ and the relative velocity $\mathbf{v}=(v_{x^\prime}, v_{y^\prime}, v_{z^\prime})$ in the observer's $x^\prime y^\prime z^\prime$ frame may be related by six mathematical relations to six parameters, i.e., $a$ (semi-major axis), $e$ (eccentricity), $\phi_0$ (argument of the periastron), and $\phi$ (phase for which the true anomaly $\Delta\phi$ is given by $\Delta\phi=\phi-\phi_0$), and two orientation parameters $i$ (inclination) and $\theta$ (rotation about the $z^\prime$ axis which is related to the argument of the ascending node $\Omega$ by $\Omega =\pi -\theta$ in the figure). Under the stated assumptions, exact (extremely precise) values of $\mathbf{r}$ and $\mathbf{v}$ would uniquely determine all six parameters, and realistic values with measurement uncertainties will allow well-determined PDFs of the parameters.  

In generalized(pseudo) Newtonian gravity, the motion of star B relative to star A (brighter) satisfies the orbit equation
\begin{equation}
    r = \frac{a(1-e^2)}{1+e\cos(\phi-\phi_0)},
    \label{eq:orbit}
\end{equation}
and the vis-viva equation
\begin{equation}
    v = \sqrt{\gamma f_M \frac{G_{\rm N} M_{\rm tot}}{r} \left(2-\frac{r}{a}\right)},
    \label{eq:visviva}
\end{equation}
where $M_{\rm tot} (\equiv M_A + M_B)$ is the observationally estimated total mass, $\gamma=G/G_{\rm N}$, and $f_M$ is a factor to take into account the uncertainty of the observational mass. 

The six observables of $\mathbf{r}$ and $\mathbf{v}$ are related to the six free parameters ($a$, $e$, $i$, $\phi_0$, $\phi$, and $\theta$) by the following six mathematical relations (see Appendix~A of \cite{Chae:2025}):
\begin{equation}
    \left[\begin{array}{c} b (\equiv y^\prime/x^\prime) \\ s \left(\equiv \sqrt{{x^\prime}^2+{y^\prime}^2}\right) \\ z^\prime   \end{array} \right] = \left[\begin{array}{c}
       \frac{-\tan\theta+\tan\phi\cos i}{1 + \tan\phi\tan\theta \cos i}  \\
         r \sqrt{\cos^2\phi+\sin^2\phi\cos^2 i} \\
         -r\sin\phi\sin i \\
    \end{array} \right]
    \label{eq:relpos_model},  
\end{equation}
and
\begin{equation}
    \left[\begin{array}{c} v_{x^\prime} \\ v_{y^\prime}  \\ v_{z^\prime}   \end{array} \right] = k v  |\cos\psi|\left[\begin{array}{c}
        \cos\theta + \tan\psi  \cos i \sin\theta  \\
       -\sin\theta + \tan\psi \cos i \cos\theta  \\
       -\tan\psi \sin i   \\
    \end{array} \right],
    \label{eq:relvel_model}  
\end{equation}
where $r$ and $v$ are given by Equations~(\ref{eq:orbit}) and (\ref{eq:visviva}), $k(=\pm 1)$ indicates the sign of the quantity $dx/d\phi$ or $\sin\phi + e\sin\phi_0$, and the variable $\psi$ satisfies
\begin{equation}
    \tan\psi = - \frac{\cos\phi + e\cos\phi_0}{\sin\phi + e\sin\phi_0}.
    \label{eq:psi}
\end{equation}
In Equation~(\ref{eq:relpos_model}), two variables $b=b(x^\prime,y^\prime)$ and $s=s(x^\prime,y^\prime)$ are used instead of $x^\prime$ and $y^\prime$ themselves for reasons to be clear soon.

\emph{Any} inference of the orbit and orientation parameters is based on the above six relations or equivalent ones. For ideal data with negligible errors for all six quantities of $\mathbf{r}$ and $\mathbf{v}$, the above relations provide the equations that uniquely determine all six parameters and no statistical analysis is necessary. For real data with uncertainties, the only issue is how to implement the six relations. Here, it is important to note that if any of the six quantities in the left side of Equations~(\ref{eq:relpos_model}) and (\ref{eq:relvel_model}) is(are) essentially exact (i.e.\ extremely precise with negligible error(s)), the relation(s) including that quantity(-ies) provide(s) the mathematical constraint(s) (i.e., consistency relation(s)) for the involved parameters, and the quantity(-ies) need(s) not be included in a likelihood of the fitting problem. This means that the mathematical constraint(s) reduce the number(s) of free parameters because one parameter per constraint is automatically determined from the other parameters. 

Let $\alpha_I$ and $\delta_I$ denote the measured R.A. and the decl.\ of star $I$ ($I=A,B$ hereafter) in units of degrees and $d_I$ denote the measured distance of star $I$ in units of parsec(pc), then we have the observed ratio
\begin{equation}
  b_{\rm obs} =  - \frac{\Delta \delta}{\Delta\alpha}\left[\cos\left(\frac{\delta_A+\delta_B}{360/\pi}\right) \right]^{-1},
\label{eq:b_obs}
\end{equation}
and the observed sky-projected separation of $B$ from $A$ in the plane defined in Figure~\ref{fig:Euler2angles}
\begin{equation}
  s_{\rm obs} =  \frac{18}{5}d_A\sqrt{\cos^2\left(\frac{\delta_A+\delta_B}{360/\pi}\right) (\Delta\alpha)^2+(\Delta\delta)^2}\text{ kau},
\label{eq:s_obs}
\end{equation}
where $\Delta\alpha\equiv \alpha_B-\alpha_A$ and $\Delta\delta \equiv \delta_B-\delta_A$. In Equation~(\ref{eq:s_obs}), we use $d_A$ (not the error-weighted mean of $d_A$ and $d_B$ as in \cite{Chae:2026}) because the actual measurement uncertainty of $s_{\rm obs}$ will be considered. Since the Gaia measured $\Delta\alpha$ and $\Delta\delta$ have negligible errors, $b_{\rm obs}$ can be treated as a constant and the error of $s_{\rm obs}$ is due solely to the error of $d_A$. With $b_{\rm obs}$ the parameter $\theta$ is a variable dependent on the other fitted parameters through 
\begin{equation}
    \tan\theta = \frac{\tan\phi \cos i - b_{\rm obs}}{1 + b_{\rm obs}\tan\phi \cos i}.
    \label{eq:theta}
\end{equation}

Now for wide binaries under consideration in 3D orbit modeling \citep{Chaeetal:2026}, $d_A$ from Gaia DR3 has a typical error of $\la 0.1\%$. Since this fractional error is at least an order-of-magnitude smaller than those for $z^\prime$ and the velocity components, one may treat it as a constant: this is the approach taken by \cite{Chae:2026}. However, in this work $s_{\rm obs}$ along with its uncertainty will be treated as an observational constraint rather than a constant. In doing so, the semi-major axis $a$ is treated as an independent variable.

The other observational quantities are given as follows:
\begin{equation}
   z^\prime_{\rm obs} = -(d_B - d_A)648/\pi \text{ kau}, 
\label{eq:zprime_obs}
\end{equation}
and
\begin{equation}
\left. \begin{array}{rl}
  v_{x^\prime,{\rm obs}} = & -4.7404 d_M(\mu^\star_{\alpha,B}-\mu^\star_{\alpha,A})\text{ m\,s}^{-1} \\ 
  v_{y^\prime,{\rm obs}} = & 4.7404 d_M(\mu_{\delta,B} -\mu_{\delta,A})\text{ m\,s}^{-1} \\
  v_{z^\prime,{\rm obs}} = & -1000(\text{RV}_B - \text{RV}_A) \text{ m\,s}^{-1}
\end{array} \right\},
\label{eq:relvel_obs}
\end{equation}
where $\mu^\star_{\alpha,I}$ and $\mu_{\delta,I}$ are PMs in the R.A. and decl.\ directions given in units of of mas~yr$^{-1}$, and RV$_I$ are the measured RVs given in units of km\,s$^{-1}$. Here, PMs and RVs have been corrected for the spherical geometric (perspective) effects using the convenient prescriptions given by \cite{Yoon:2025}, and thus error-weighted mean distances ($d_{\rm M}$) are used to derive $v_{x^\prime,{\rm obs}}$ and $v_{y^\prime,{\rm obs}}$ for the perspective effect-corrected PMs. 

The Bayesian approach of \cite{Chae:2026} is followed with the adjustments introduced above. The observed quantities to constrain the model parameters are $\mathbf{X}_{\rm obs} = \{s_{\rm obs}, z^\prime_{\rm obs}, v_{x^\prime,{\rm obs}}, v_{y^\prime,{\rm obs}}, v_{z^\prime,{\rm obs}}$\}, and their model predictions ($\mathbf{X}_{\rm mod}$) are given in Equations~(\ref{eq:relpos_model}) and (\ref{eq:relvel_model}). The model free parameters for the general case are $\mathbf{\Theta}=\{a, e,i,\phi_0,\Delta\phi(\equiv\phi-\phi_0),f_M,\Gamma \}$. Only the parameter $\theta$ is a dependent variable through Equation~(\ref{eq:theta}). The posterior probability distribution of the parameters $p(\mathbf{\Theta})$ is given by
\begin{equation}
    \ln p(\mathbf{\Theta}) = \ln\mathcal{L} + \sum_l \ln f_\text{pr}(\Theta_l),
\label{eq:postprob}    
\end{equation}
where $\mathcal{L}$ is the likelihood function, and $f_\text{pr}$($\Theta_l$) ($l=1,\cdots,7$) is the prior probability density function for the parameter $\Theta_l$. The likelihood function is defined by
\begin{equation}
    \ln\mathcal{L} = -\frac{1}{2}\sum_j \left[ \left(\frac{X_{\rm{mod},j}(\mathbf{\Theta})-X_{\rm{obs},j}}{\sigma_j} \right)^2 +\ln(2\pi\sigma_j^2) \right],
\label{eq:likelihood}    
\end{equation}
where $\{X_{\rm{mod},j}(\mathbf{\Theta})\}$ ($j=1,\cdots,5$) are the model predictions while $X_{\rm{obs},j}$ and $\sigma_j$ are the observed quantities and their uncertainties. The physical ranges of the angular variables are $0\le i \le \pi$, $0\le \phi_0<2\pi$, $0\le \Delta\phi <2\pi$, and $-\pi\le\theta<\pi$. However, the calculation is actually performed for $-\pi\le i \le \pi$ and $-\pi/2<\theta<\pi/2$ (the range allowed by Equation~(\ref{eq:theta})) using mathematical transformations to the physical ranges described in \cite{Chae:2026}.

The priors on the parameters are given as follows: $f_\text{pr}(e) = 2e$ (the thermal probability distribution), $f_\text{pr}$($i$) = $|\sin(i)|$, $f_\text{pr}$($\phi_0$) = uniform, and 
\begin{equation}
    f_\text{pr}(\Delta\phi) = \frac{(1-e^2)^{3/2}}{2\pi} \frac{1}{[ 1+ e \cos(\Delta\phi) ]^2}. 
\label{eq:PrDelphi}   
\end{equation}
For $f_M$, a lognormal probability distribution with mean $=0$ and s.d.\ $=5\%$ is assumed. When gravity is free, a flat prior is assumed for $\Gamma$ in the range $-1<\Gamma<1$. Finally, for the semi-major axis $a$, the following prior is considered for $A\equiv\ln(a/{\rm kau})$:
\begin{equation}
 f_{\rm pr}(A) = c\left\{ \begin{array}{ll}
  \exp(-0.6A_{\rm break}) & \text{for }A\le A_{\rm nreak} \\ 
  \exp(-0.6A)             & \text{for }A> A_{\rm nreak}
\end{array} \right.,
\label{eq:PrA}     
\end{equation}
where $A_{\rm break}=\ln(5)$ and $c$ is the normalization factor determined for the range $(A_{\rm min},A_{\rm max})$ with $A_{\rm min}\equiv -1+\ln(s_{\rm obs}/{\rm kau})$ and $A_{\rm max}\equiv2.5+\ln(s_{\rm obs}/{\rm kau})$. A prior equivalent to the prior of Equation~(\ref{eq:PrA}) was introduced by \cite{SaadTing:2026} based on \cite{Andrews:2017}. This prior will be considered only to test its possible effects in Bayesian inferences. In testing the effects of treating $a$ as a free parameter, the prior of Equation~(\ref{eq:PrA}) will not be implemented for a consistent comparison with the results based on the algorithm of \cite{Chae:2026}.

The numerical procedure to derive the posterior distribution $p(\mathbf{\Theta})$ is the same as described in \cite{Chae:2026}: Markov chain Monte Carlo (MCMC) procedure is performed with the public Python package {\tt emcee} \citep{emcee}.

\section{Results with the Extended Algorithm and Comparison with Other Results} \label{sec:result}

For the purpose of testing the effects of how the semi-major axis $a$ is treated, we consider binary systems for which detailed orbit solutions are available from independent studies. First, we consider the binary of stars HD~189739 and HD~189760 (Binary {\#}24 from \cite{Saglia:2025}). This is one of the systems from the recent compilation by \cite{Chaeetal:2026} that does not allow bound Newtonian orbits and indicates $\Gamma>0$ based on the \cite{Chae:2026} algorithm that uses the sky-projected separation $s_{\rm obs}$ as a numerical constant and treats $a$ as a dependent variable. It is interesting whether the extended algorithm can modify the \cite{Chae:2026} inference in a meaningful way. Second, we consider the $\alpha$-Cen AB - Proxima system that has exceptional-quality data and thus provides an ideal system to compare results with various algorithms.

\begin{figure*}[!htb]
    \centering
    \includegraphics[width=1.\linewidth]{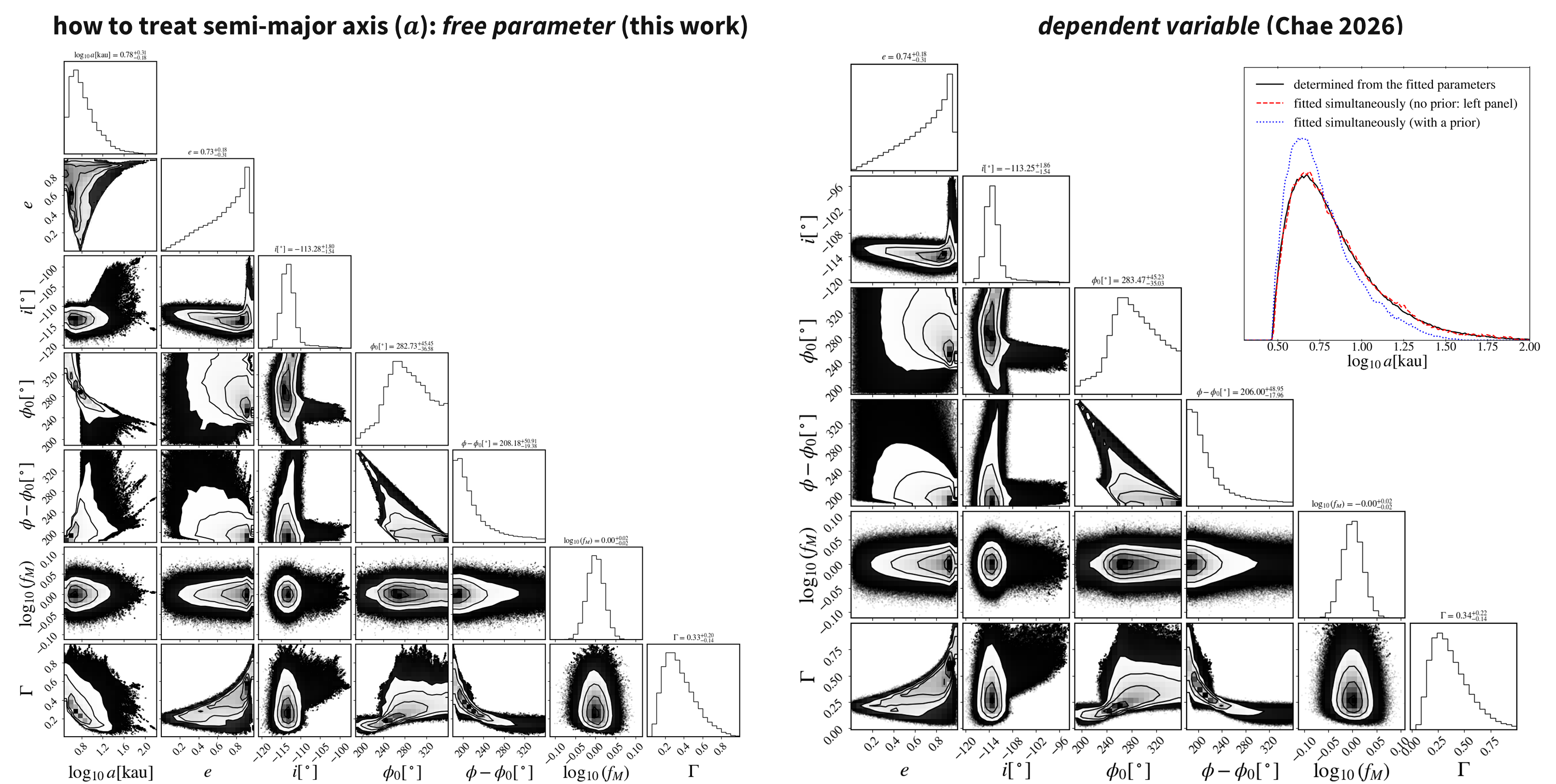}
    \caption{
    This figure shows posterior PDFs of parameters for the binary of HD~189739 and HD~189760. The left panel is based on the extended algorithm of this work in which the semi-major axis ($a$) is fitted simultaneously with the other parameters, while the right panel is based on the \cite{Chae:2026} algorithm in which $a$ is not a free parameter but a variable dependent on the other parameters through a mathematical constraint. The inset compares the distributions of $a$ in both cases, which agree well with each other. The dotted curve in the inset shows the distribution of $a$ with the prior of Equation~(\ref{eq:PrA}) imposed.
    }
    \label{fig:PDFexample_afree}
\end{figure*}

The left panel of Figure~\ref{fig:PDFexample_afree} shows posterior PDFs of the parameters in the extended algorithm for the binary of HD 189739 and HD 189760. The right panel shows the PDFs of the parameters in the \cite{Chae:2026} algorithm. In that algorithm, $a$ is a dependent variable that is determined by
\begin{equation}
    a = s_{\rm obs}\frac{1+e\cos(\phi-\phi_0)}{(1-e^2)\sqrt{\cos^2\phi+\cos^2i\sin^2\phi}}
    \label{eq:a_semimajor}
\end{equation}
for the observed $s_{\rm obs}$ (ignoring the tiny measurement error) and the other fitted parameters. The PDFs of the fitted parameters with the extended algorithm are consistent with those of the \cite{Chae:2026} algorithm. Also, as the inset in the right panel shows, the distribution of directly fitted $a$ agrees well with that inferred with Equation~(\ref{eq:a_semimajor}) for the parameters fitted with the \cite{Chae:2026} algorithm. If the prior given by Equation~(\ref{eq:PrA}) is imposed on $a$ in using the extended algorithm, the inferred distribution is mildly changed as shown by the dotted curve.

\begin{figure}[!htb]
    \centering
    \includegraphics[width=1.\linewidth]{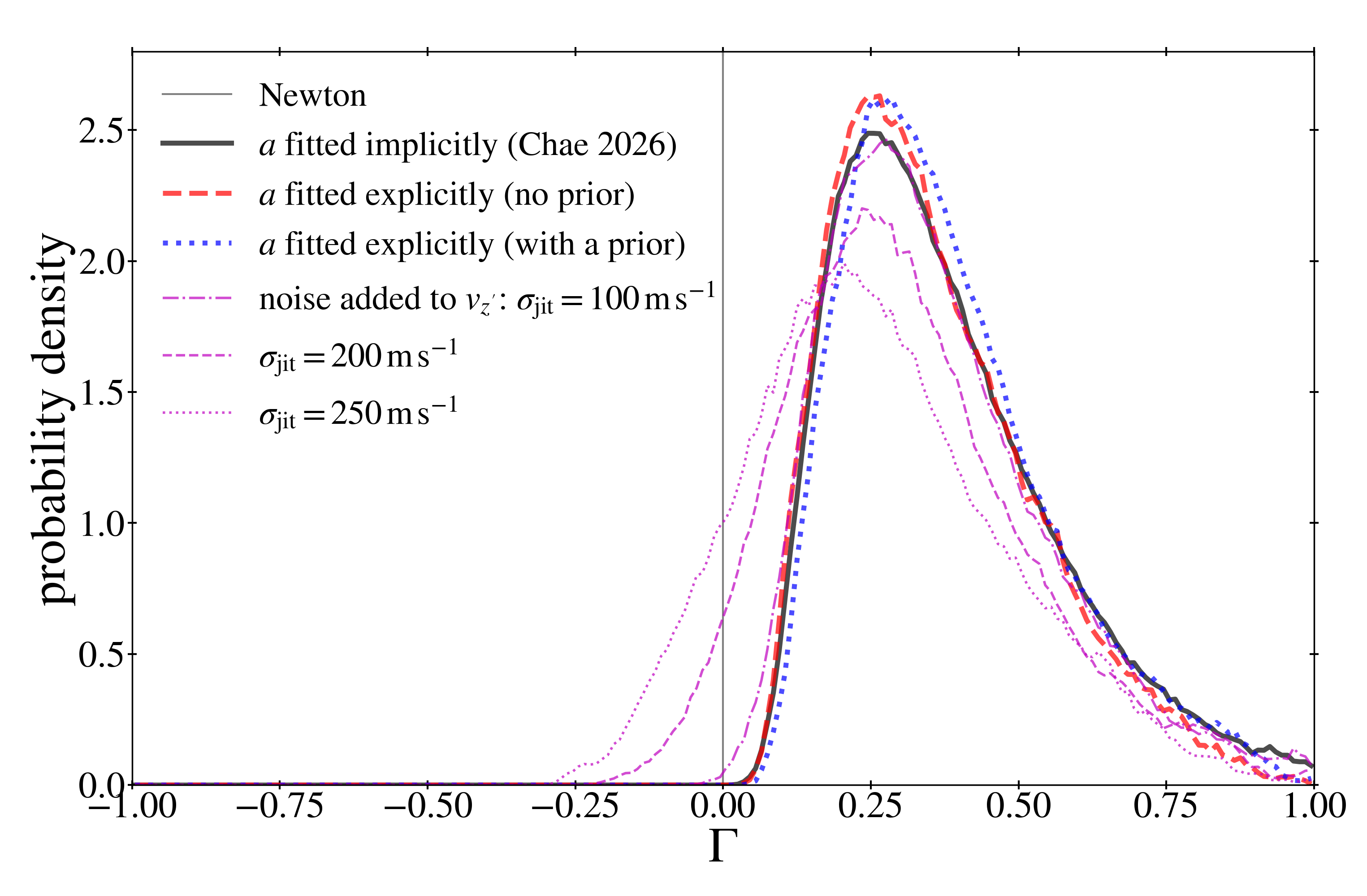}
    \caption{
    This figure shows posterior PDFs of $\Gamma$ from various runs of Bayesian MCMC modeling for the binary of HD~189739 and HD~189760. With the given data \citep{Saglia:2025}, Newton is completely excluded whether the semi-major axis is fitted implicitly \citep{Chae:2026} or explicitly with or without the prior of Equation~(\ref{eq:PrA}). Newton would be consistent with this system only if some data had much larger errors than reported. Here cases of adding an arbitrary jitter error ($\sigma_{\rm jit}$) to the relative radial velocity are shown as examples of easing tension with Newton. 
    }
    \label{fig:Gamma_distribution}
\end{figure}

\begin{figure*}[!htb]
    \centering
    \includegraphics[width=1.\linewidth]{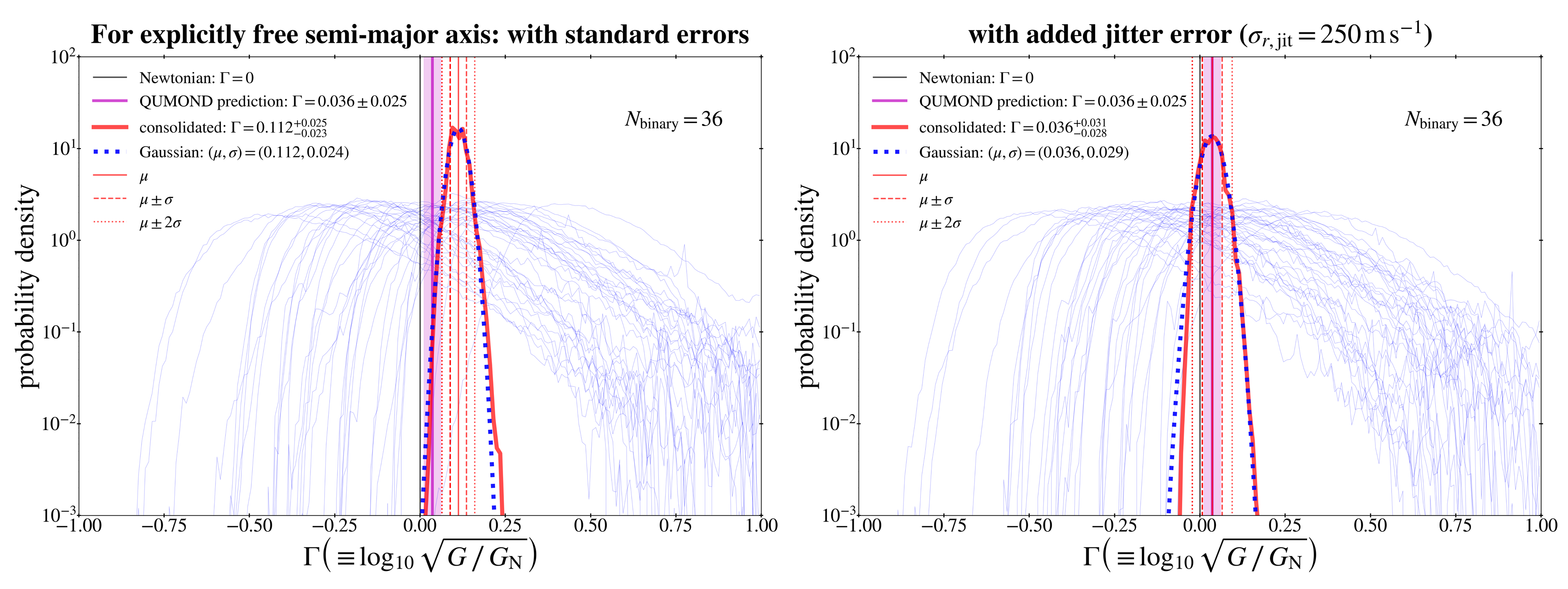}
    \caption{
    This figure shows posterior PDFs of $\Gamma$ for all 36 wide binaries from the `clean' sample by \cite{Chaeetal:2026}. The consolidated PDF of the individual PDFs and a QUMOND prediction are also displayed. The left panel shows the result with the measurement errors as given by observations. The right panel shows a hypothetical result with an arbitrary jitter error of $250\,{\rm m}\,{\rm s}^{-1}$ uniformly added to the relative radial velocities $v_r(=v_{z^\prime})$ of all wide binaries.
    }
    \label{fig:Gamma_afree}
\end{figure*}

Because all parameters are equivalently constrained whether $a$ is fitted explicitly or implicitly, the inference of $\Gamma$ is also unaffected, as shown in Figure~\ref{fig:Gamma_distribution}. Newton is completely inconsistent with this system because the PDF declines more sharply than a Gaussian curve (the nominal 68\% confidence interval does not correctly capture the whole PDF). Thus, as long as systems such as this are not excluded by a prior choice, the \cite{Chae:2026} sample as a whole cannot agree with Newton simply by treating $a$ as an explicitly free parameter, as shown in the left panel of Figure~\ref{fig:Gamma_afree}. The consolidated value of $\Gamma=0.112_{-0.023}^{+0.025}$ is fully consistent with the value of $\Gamma=0.102_{-0.021}^{+0.028}$ obtained by \cite{Chaeetal:2026} with the \cite{Chae:2026} algorithm in which $a$ is implicitly determined.

In contrast to the above results, which unambiguously prove that whether fitting $a$ implicitly or explicitly is not an issue, \cite{SaadTing:2026} claimed that $\gamma\approx1.0$ resulted for the \cite{Chaeetal:2026} sample in their baseline model where $a$ is an explicitly free parameter. Then, the question is how they got such a result, as treating $a$ as a free parameter cannot be the cause. In their paper, they state that they correctly use a median error of $0.1\%$ for the projected separation $s$ (the last paragraph of their Section~2). We notice that their modeling introduces jitter errors for the relative velocity components whose values are not specified in their paper. Taking their words in their paper, we first perform some numerical exercises to investigate possible effects of arbitrary jitter errors for the relative radial velocity which has the largest observational uncertainty\footnote{The nominal uncertainties of radial velocities are extremely small for systems such as the binary of HD 189739 and HD 189760, but the uncertainty of correcting for gravitational redshift and convective blueshift is estimated to be $40\,{\rm m}\,{\rm s}^{-1}$ \citep{Saglia:2025}.}: $\sigma_{\rm jit}=100\,{\rm m}\,{\rm s}^{-1}$, $200\,{\rm m}\,{\rm s}^{-1}$, and $250\,{\rm m}\,{\rm s}^{-1}$, which are 2.5, 5.0, and 6.25 times as large as the estimated error. For $\sigma_{\rm jit}=100\,{\rm m}\,{\rm s}^{-1}$, Newton is still inconsistent, but tension with Newton is significantly lessened with $\sigma_{\rm jit}\ga 200\,{\rm m}\,{\rm s}^{-1}$. In cases like the latter one, the \cite{Chaeetal:2026} sample would be consistent with Newton, as shown in the right panel of Figure~\ref{fig:Gamma_afree}, given that this binary is one of the most anomalous systems based on the reported errors. However, errors of $\ga 200\,$ for this system are not viable based on the available data. 

\begin{figure}[!htb]
    \centering
    \includegraphics[width=1.\linewidth]{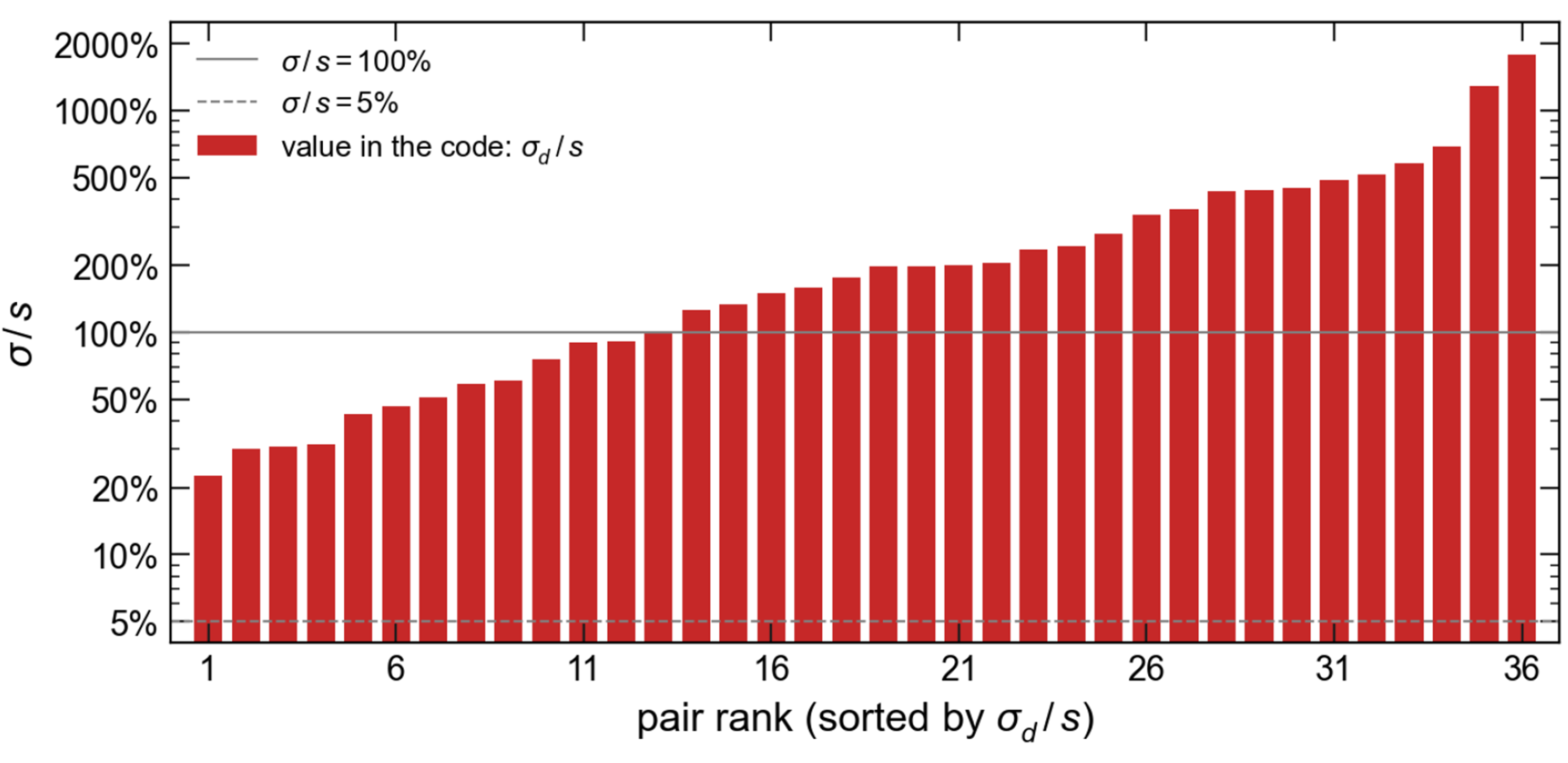}
    \caption{
    This figure shows the distribution of $\sigma_d/s$ in 36 wide binaries used by \cite{SaadTing:2026}. For most wide binaries, $\sigma_d$ is the actual error for $s$ used in the baseline model based on the open code.
    }
    \label{fig:derr_rp}
\end{figure}

Given that large uniform jitter errors for $v_r$ are not viable for the \cite{Chaeetal:2026} sample, we have checked their code itself\footnote{https://github.com/seratsaad/wb3d-gamma} and noticed a fatal error in their implementation of $s$. They set the error of their $s$ as the maximum of $0.05\times s$ and $\sigma_d \left(\equiv 0.5\times \sqrt{\sigma_{d_A}^2+\sigma_{d_B}^2}\right)$.\footnote{Please see lines 105 and 106 of {\tt run\_gamma\_analysis.py}.} The latter is almost always larger than the former and of the same order as $s$ or even much larger in some cases, as shown in Figure~\ref{fig:derr_rp}. This does not agree with a median error of $0.1\%$ that they quote in their paper. The large errors actually used by their `baseline model' mean that their orbits are essentially independent of the measured velocity components used in the modeling. Their orbits are statistically assigned using their prior on $a$ (Equation~(\ref{eq:PrA})) irrespective of the actually observed orbit scale. The agreement of the baseline model results with Newton is not a valid or meaningful result but a consequence of this effective distortion of the data.

\begin{figure*}[!htb]
    \centering
    \includegraphics[width=1.\linewidth]{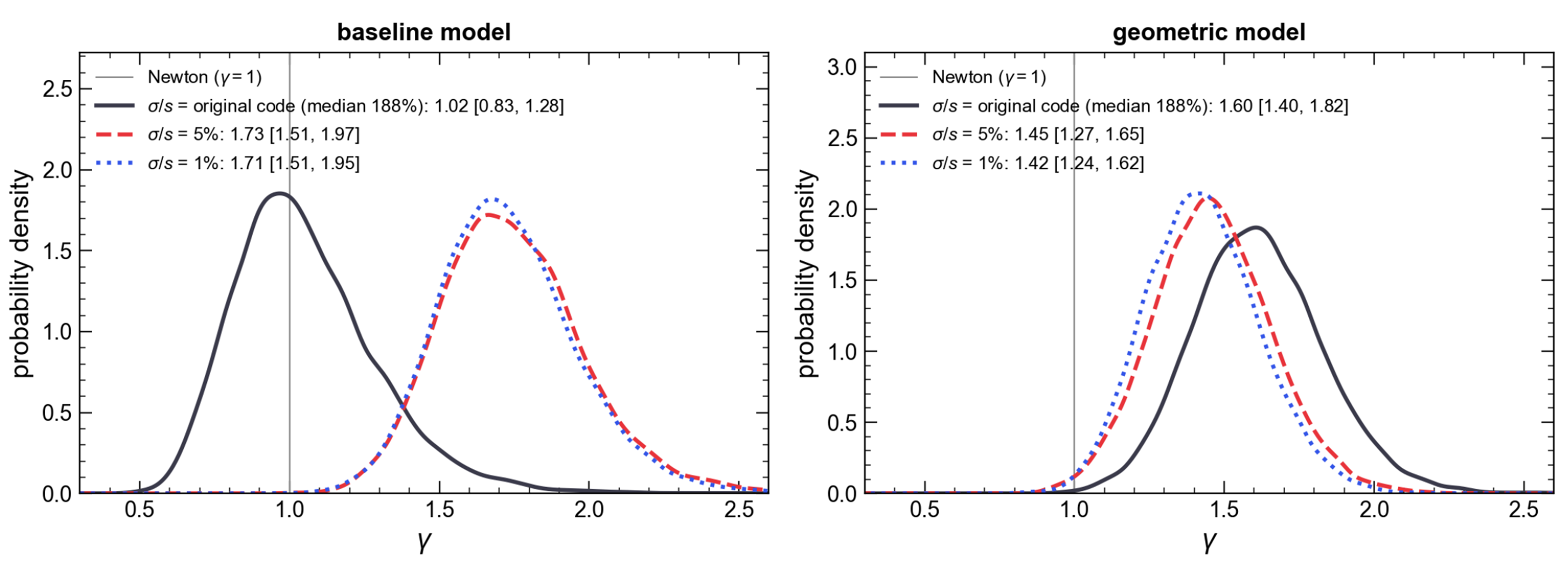}
    \caption{
    This figure compares original results by \cite{SaadTing:2026} based on extremely large errors of $s$ (as shown in Figure~\ref{fig:derr_rp}) with the results produced in this work with the same code but based on much smaller errors of $5\%$ or $1\%$.
    }
    \label{fig:SaadTing_corrected}
\end{figure*}

Then, since their geometric (deprojection) model (in which the observed $s$ is used as a self-consistency relation) returns a correct result agreeing with that by \cite{Chaeetal:2026}, their baseline model with correct errors for $s$ should agree with the geometric model result if all the other calculations are consistent. We therefore recalculated the baseline model by replacing the incorrect errors of $s$ with much smaller errors of 5\% or 1\%.\footnote{We have also tried a 0.1\% error for $s$, but the numerical procedure did not converge well; however, 5\% or 1\% error for $s$ seems to constrain the orbit sufficiently well.} The results are shown in Figure~\ref{fig:SaadTing_corrected}. The agreement between the baseline and the geometric results is much better now than in the original results by \cite{SaadTing:2026}. However, the agreement is not as good as that between the $a$-implicit result based on the \cite{Chae:2026} algorithm and the $a$-explicit result based on the extended algorithm presented here. It is unclear where the difference comes from, as the difference between the baseline and the geometric results should disappear as the error of $s$ gets smaller and smaller. Despite the remaining small difference, this numerical exercise clearly shows that the absurdly large errors of $s$ used by the baseline model were indeed responsible for the Newtonian result by \cite{SaadTing:2026}.

All the above tests are based on the six mathematical relations given by Equations~(\ref{eq:relpos_model}) and (\ref{eq:relvel_model}) between the six observables and the six parameters, and the tests are mainly concerned with possible effects of how to fit the semi-major axis $a$. These equations represent a possible formulation (parameterization) of the same astrodynamics that is assumed to be Keplerian with generalized $G$. It is of interest to further test the extended algorithm presented in this work with a completely independent orbit calculation that does not use the parametrization of this work. For this purpose, the $\alpha$-Cen AB - Proxima system is ideal as they provide exceptional data for which an independent orbit solution is available \citep{Kervella:2017}. Although updated (perhaps more reliable) data may be available, we will use only the data presented in \cite{Kervella:2017} so that the results with the algorithm of this work can be consistently compared with those of \cite{Kervella:2017}.

\begin{figure*}[!htb]
    \centering
    \includegraphics[width=1.\linewidth]{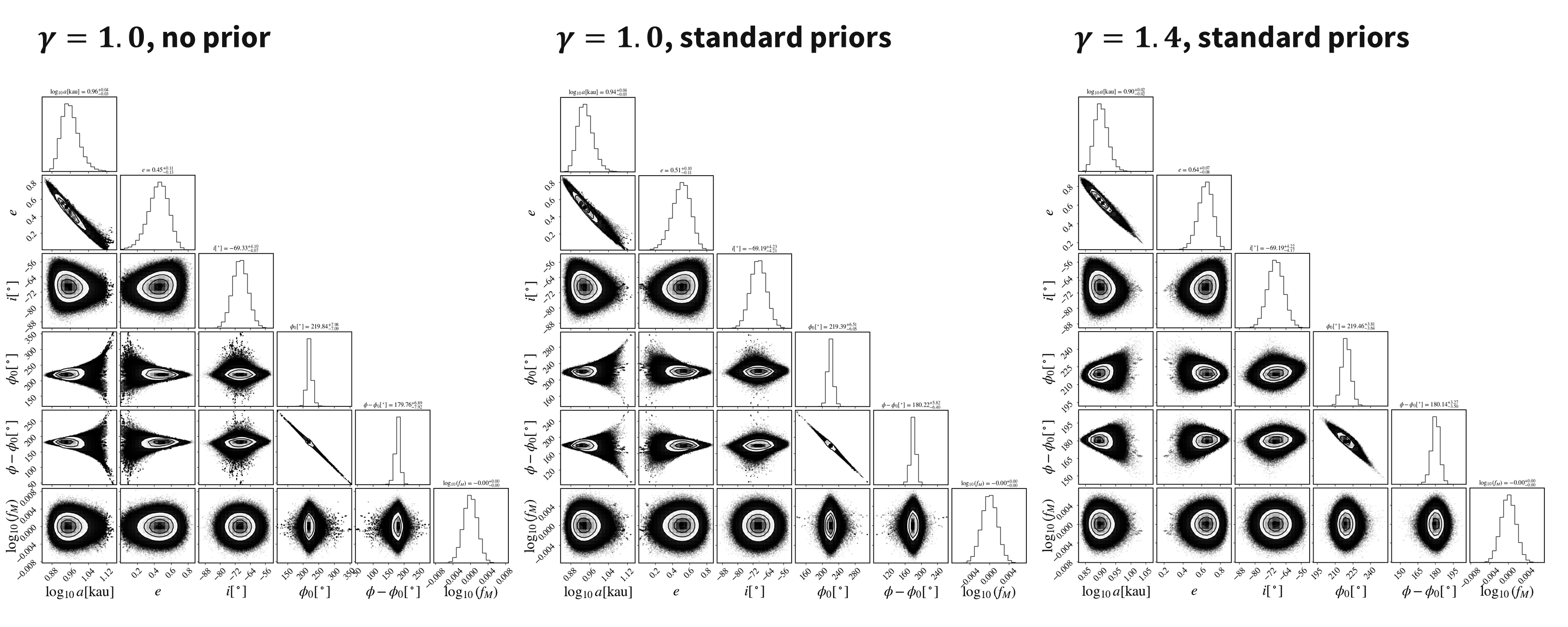}
    \caption{
    This figure exhibits obit solutions in fixed $G=\gamma G_{\rm N}$ for Proxima \citep{Kervella:2017} with the algorithm of this work. The left panel presents the Newtonian solution without any prior on any of the six orbit and orientation parameters. The middle panel shows the Newtonian solution with the standard priors of this work (i.e., all the priors except for the prior of Equation~(\ref{eq:PrA}) on $a$). The right panel shows the solution with the standard priors in boosted gravity with $\gamma=1.4$.  
    }
    \label{fig:PDFs_Proxima}
\end{figure*} 

Because the effects of the spherical geometry are important for the large angular separation of $\approx 2.2^\circ$ between $\alpha$-Cen AB (barycenter) and Proxima, Equation~(\ref{eq:relvel_obs}) cannot be used to estimate relative velocity components from measured PMs and RVs. \cite{Kervella:2017} obtained the components of the relative displacement and velocity between $\alpha$-Cen AB and Proxima in the Galactic frame and obtained their orbit solutions in that frame. In the algorithms of \cite{Chae:2026} and this work, the (local) Cartesian framework defined by Figure~\ref{fig:Euler2angles} is used. The components of $\mathbf{v}$ in the $x^\prime y^\prime z^\prime$ frame may be computed taking into account the effects of the spherical geometry using the formulae given in \cite{Yoon:2025}. Although the precision of the derived components of $\mathbf{v}$ based on the \cite{Yoon:2025} formulae is adequate for the test, we find that the coordinate transformations described in Appendix~\ref{sec:coordtranf} provide slightly more accurate components of $\mathbf{v}$ whose transformed components in the Galactic frame match the relative velocity components obtained by \cite{Kervella:2017}. 

Since \cite{Kervella:2017} presented orbit solutions in Newtonian gravity, we first obtain solutions in Newtonian gravity for the purpose of comparing independent statistical inferences for the same data. \cite{Kervella:2017} and this work are different in three aspects. First, \cite{Kervella:2017} worked in the Galactic frame while this work uses the local Cartesian frame defined by Figure~\ref{fig:Euler2angles}. Second, \cite{Kervella:2017} used a classical Monte Carlo approach in which $10^5$ sets of $\{M_{\rm tot},\mathbf{r},\mathbf{v}\}$ consistent with random fluctuations were used to derive distributions of the parameters, while this work employs the Bayesian MCMC approach and uses $2\times 10^6$ sets of models to derive the PDFs of the parameters. Finally, this work tests cases with or without priors on the parameters while the classical approach does not incorporate priors.  

As in the binary of HD~189739 and HD~189760, the explicit-$a$ algorithm and the implicit-$a$ algorithm give completely equivalent results for the $\alpha$-Cen - Proxima system given the same priors. Thus, we show only the explicit-$a$ results. The left panel of Figure~\ref{fig:PDFs_Proxima} shows the PDFs of the parameters in Newtonian gravity without any priors on the orbit and orientation parameters, while the middle panel shows the results with the standard priors of this work. The no-prior case can be more directly compared with the \cite{Kervella:2017} results. Although this work and \cite{Kervella:2017} use different frames, we can test whether orientation-independent parameters, i.e., eccentricity ($e$), semi-major axis ($a$), and orbit true anomaly ($\Delta\phi$) agree between the two results. 

\cite{Kervella:2017} obtained $e=0.50_{-0.09}^{+0.08}$, $a=8.7_{-0.4}^{+0.7}\,{\rm kau}$, and $\Delta\phi = 177.6_{-15.5}^{+10.4}$\,degrees (estimated from their values of period and `epoch of periastron'). The no-prior output from this work gives $e=0.45_{-0.13}^{+0.11}$, $a=9.1_{-0.6}^{+0.9}\,{\rm kau}$, and $\Delta\phi= 179.8_{-7.9}^{+6.9}$\,degrees. The two results agree within the estimated uncertainties. However, there are some qualitative differences regarding the estimated uncertainties. The MCMC results have somewhat larger uncertainties for $e$ and $a$ but smaller uncertainties for $\Delta\phi$. In particular, the uncertainties of $\Delta\phi$ from the classical MC procedure are $\approx 2$ times as large as those from the MCMC procedure. Uncertainties of orbits can be represented by any parameters but the classical MC procedure tends to assign uncertainties more heavily on $\Delta\phi$ than the MCMC procedure. This shows that how models are sampled (i.e., independent or Markov chain) can somewhat matter for the present problem, although the two results are statistically well consistent. With the standard priors (the middle panel of Figure~\ref{fig:PDFs_Proxima}), the MCMC procedure returns $e=0.51_{-0.11}^{+0.10}$, $a=8.7_{-0.6}^{+0.8}\,{\rm kau}$, and $\Delta\phi= 180.2_{-6.4}^{+5.8}$\,degrees. Interestingly, the values of $e$ and $a$ with the priors better match the values obtained by \cite{Kervella:2017}.

It is of interest how the inference of orbits is affected if gravity is modified from Newton since the mutual internal Newtonian acceleration is in the MOND \citep{Milgrom:1983} regime. The right panel of Figure~\ref{fig:PDFs_Proxima} shows the results for $\gamma=1.4$ which is a typical boost factor predicted by MOND gravity \citep{BanikZhao:2018,ChaeMilgrom:2022,Pf-A:2025}: $e=0.64_{-0.08}^{+0.07}$, $a=7.9_{-0.5}^{+0.6}\,{\rm kau}$, and $\Delta\phi= 180.1_{-3.5}^{+3.3}$\,degrees. Although not shown in a figure, the results with $\gamma=1.255$ (which corresponds to the specific prediction of \cite{Pf-A:2025}) are $e=0.60_{-0.09}^{+0.08}$, $a=8.1_{-0.4}^{+0.6}\,{\rm kau}$, and $\Delta\phi= 180.2_{-4.2}^{+3.9}$\,degrees. Thus, MONDian boosted gravity predict somewhat more eccentric orbits. Here, it is interesting to note that wide binaries of large separations like the present system statistically prefer more eccentric orbits than $e\approx 0.5$ \citep{hwang:2022}. A robust prediction of the Bayesian MCMC procedure in all cases of $\gamma$ considered is that Proxima is almost right at the apastron phase. The \cite{Kervella:2017} MC procedure also predicts such a phase but with much larger uncertainties.

\section{Discussion and Conclusions} \label{sec:conc}

Wide binary gravity tests have been maturing to the point that accurate 3D relative velocities between the two stars have been measured and 3D orbit modeling has been performed, resulting in discoveries of individual anomalous systems \citep{Saglia:2025,Chaeetal:2026} and statistical inferences of the effective gravitational parameter $G$ or the boost factor $\gamma$ (or $\Gamma$) \citep{Chae:2026,Chaeetal:2026}. While anomalous systems can be detected by comparing the magnitude of the measured 3D relative velocity $v$ with an upper bound on the Newtonian escape velocity even without any detailed 3D orbit modeling, a statistical inference of $\gamma$ in a sample relies on a correct methodology.

In light of the confusion in the literature \citep{SaadTing:2026}, we have carefully examined all key issues of 3D modeling and statistical inference. A 3D modeling of a wide binary system refers to fitting six parameters (four orbit parameters in the orbit plane and two orientation parameters with respect to the observer) to six observable components of $\mathbf{r}$ and $\mathbf{v}$ (in practice, the total mass of the system is also a free parameter and the measured mass provides the corresponding observable). Once Keplerian orbits are assumed (a completely valid assumption to test Newtonian gravity), the six equations (i.e., the mathematics) are fixed. Various coordinate frames and parameterization can be used for the same mathematics. There are several important points that have been addressed in this work.

First, it does not matter whether an equation involving extremely precise observable(s) is used as a fixed internal mathematical relation to be satisfied by relevant parameters or is used as an explicit observational constraint with the measured uncertainty. This is explicitly tested in this work regarding how to use the observed sky-plane separation $s_{\rm obs}$ between the two stars. In the first option, the fixed equation (Equation~(\ref{eq:a_semimajor})) can be used as an internal relation to make the semi-major axis $a$ a variable dependent on the other parameters, while in the second option, it is explicitly used as an observable to be included in the likelihood (Equation~(\ref{eq:likelihood})), so that $a$ is treated explicitly free and fitted simultaneously together with the other parameters. The two options give equivalent Bayesian results, as demonstrated in Figure~\ref{fig:PDFexample_afree} and Figure~\ref{fig:Gamma_afree}. The claim by \cite{SaadTing:2026} that treating $a$ explicitly free can change the Bayesian inference dramatically is invalid and \emph{logically impossible}.

Second, a relation in Equation~(\ref{eq:relpos_model}) given by $s=r\sqrt{\cos^2\phi+\sin^2\phi\cos^2i}$ (or an equivalent equation in any other parameterization) is one of the six fundamental equations that \emph{must} be used to properly constrain the parameters. \cite{SaadTing:2026} describes it as a deprojection\footnote{The concept of deprojection does not apply here as all six parameters are directly fitted to the six observables. The concept of deprojection was introduced by \cite{Chae:2023} when the sky-plane 2D vectors were statistically deprojected into the 3D space.} equation in a misleading way (as if it is an arbitrary or special assumption) when $s_{\rm obs}$ is used in an internal mathematical relation (assuming that the observational uncertainties of $\la 0.1\%$ can be set to zero). Whether it is used as an input to the likelihood (their baseline model) or as the internal mathematical relation should not make any tangible difference as long as the actual small uncertainties are used, as verified in this work in which the equation is included in the likelihood.

Third, statistical inferences depend critically on the uncertainties of the measured quantities. Arbitrarily enlarging or lowering reported uncertainties (particularly based on observations such as Gaia's) would be a bad practice. \cite{SaadTing:2026} obtained a Newtonian result through their baseline model for the \cite{Chaeetal:2026} sample. How to treat $a$ or how to use $s_{\rm obs}$ cannot be responsible for such a result as demonstrated in this work. We have then found that their result of $\gamma\approx 1.0$ is actually due to a fatal error in their code regarding the input error of the observed projected separation. Their code effectively distorted the data given by \cite{Chae:2026} to get the Newtonian result.

Fourth, it is verified that the approximate Cartesian frame centered on a star defined by Figure~\ref{fig:Euler2angles} can be used even for the $\alpha$-Cen AB - Proxima system although the angular separation is not as small as those in typical wide binaries with distances $\ga 10\,{\rm pc}$. This is feasible because the physical components of $\mathbf{v}$ in the $x^\prime y^\prime z^\prime$ frame can be accurately obtained by coordinate transformations (Appendix~\ref{sec:coordtranf}). The agreement between this work and \cite{Kervella:2017} based on the same data provides a crosscheck of two numerical procedures.

Finally, which MC procedure is used to derive PDFs of the parameters can matter for estimating uncertainties of the parameters, as illustrated for the $\alpha$-Cen AB - Proxima system. Bayesian MCMC procedure (this work) and a classical MC procedure \citep{Kervella:2017} return different estimates of uncertainties for the orbit true anomaly although the most likely values match well. 

As explicitly proven in this work, the mathematical and statistical framework described in \cite{Chae:2026} and here can be used to properly model 3D orbits and infer gravity in a wide binary sample. One can take either option of explicitly or implicitly fitting the semi-major axis depending on how the observed sky-projected separation is implemented (the two options are fully equivalent for samples of highest-quality such as that of \cite{Chaeetal:2026}). It seems that \cite{SaadTing:2026} obtained $\gamma\approx 1.0$ by effectively distorting the data of \cite{Chaeetal:2026}, but made a misleading argument about the modeling procedure. This does not mean that there are no important issues in wide binary gravity tests based on 3D data, and scientific investigations should focus on real issues such as the following (see also \cite{Chaeetal:2026}):
\begin{itemize}
    \item It is important to derive individual-specific PDFs of orbit parameters along with $\Gamma$ for three reasons. First, we can check whether a system is physically plausible with the derived parameters. This is particularly important when a system appears anomalous from a Newtonian point of view. In such a case, we can investigate modified gravity that may allow viable orbits. Second, individual inferences of $\Gamma$ (in modeling with free gravitational parameter) are needed to investigate the trend of any anomaly with the internal Newtonian acceleration, as such a trend can discriminate various modified gravity predictions (although the current sample size is not sufficient for this purpose). Finally, individual-specific results are important to understand where the signal for (or against) modified gravity comes from. A population-level result through a `black box' without properly understanding where the signal comes from may well produce an incorrect result by tweaking/distorting the modeling procedure and/or data. This is particularly a concern in a sensitive topic such as the present one, as one can easily be tempted to distort something for ``good" reason to get the ``right" result.
    \item For anomalous systems, follow-up observations such as RV monitoring and Speckle interferometric imaging need to be made to verify or disprove the anomalous nature. 
    \item For typical wide binaries with measured radial separations of limited precision (as the radial separations are $\la 0.1\%$ of the distances), the six observables of $\mathbf{r}$ and $\mathbf{v}$, unlike the $\alpha$-Cen system, cannot constrain well all the parameters without reasonable priors. Thus, Bayesian modeling may be the only reasonable numerical procedure.
\end{itemize}

In conclusion, the Bayesian general algorithm presented in \cite{Chae:2026} and extended in this work are extensively verified and can be used to properly infer orbits and $\Gamma$ in an explicit-$a$ way or more economical implicit-$a$ way. The algorithm is particularly useful in analyzing and understanding individual systems. A proper test of gravity with wide binaries requires properly understanding all individual systems.

\begin{acknowledgments}

The author thanks P. Kervella for helpful communications on the $\alpha$-Cen system. The author thanks all the collaborators of the ongoing project to measure gravity from wide binary samples based on the algorithms discussed in this work. This work was supported by the National Research Foundation of Korea (NRF) grant funded by the Korea government (MSIT) (RS-2026-25492976).

\end{acknowledgments}

\bibliographystyle{aasjournalv7.bst}
\bibliography{ms.bib}{}


\appendix

\section{Taking into account the perspective effects in the spherical coordinates through coordinate transformations} \label{sec:coordtranf}

When the angular separation between binary stars is not sufficiently small, the differences in the PMs and RVs of the two stars through Equation~(\ref{eq:relvel_obs}) do not provide physical relative velocities. The problem is most conspicuous in the binary of $\alpha$-cen AB and Proxima. Here we describe an accurate and simple method to obtain the correct relative velocities in the local Cartesian frame centered on a star (the $x^\prime y^\prime z^\prime$ frame in Figure~\ref{fig:Euler2angles}). 

Let star $I$ ($I=A,B$) have $\alpha_I$ (R.A.), $\delta_I$ (decl.), $d_I$ (distance from the Sun), $\mu^\star_{\alpha,I}$ (PM in the R.A. direction), $\mu_{\delta,I}$ (PM in the decl.\ direction), and RV$_I$ (radial velocity). Let the equatorial frame be described by the spherical polar coordinates $(r,\theta,\phi)=(d,\pi/2-\delta,\alpha)$ and the Cartesian coordinates $(X,Y,Z)$ defining the frame. The two stars have measured velocities in the spherical coordinates as $V_{r,I}={\rm RV}_I$, $V_{\theta,I}=-f\times d_I \times \mu_{\delta,I}$, and $V_{\phi,I}=f\times d_I \times \mu^\star_{\alpha,I}$ where $f$ is the conversion factor as in Equation~(\ref{eq:relvel_obs}). These velocities can be transformed to the $XYZ$ frame as
\begin{equation}
\left[\begin{array}{c} V_{X,I} \\ V_{Y,I} \\ V_{Z,I} \\  \end{array} \right]  = 
\left[\begin{array}{ccc}  \sin\theta_I\cos\phi_I  & \cos\theta_I\cos\phi_I  &  -\sin\phi_I \\
         \sin\theta_I\sin\phi_I  & \cos\theta_I\sin\phi_I  &  \cos\phi_I \\
          \cos\theta_I  & -\sin\theta_I  &  0  \\
    \end{array} \right] \left[\begin{array}{c} V_{r,I} \\ V_{\theta,I} \\ V_{\phi,I} \\  \end{array} \right] .  
    \label{eq:coordtrans}  
\end{equation}
The relative velocities in this frame are obtained by $v_X\equiv V_{X,A}-V_{X,B}$, $v_Y\equiv V_{Y,A}-V_{Y,B}$, and $v_Z\equiv V_{Z,A}-V_{Z,B}$. 

Let the physical relative velocities in the local Cartesian frame centered on star A be denoted by $v_r$, $v_\theta$, and $v_\phi$. They can be obtained by 
\begin{equation}
\left[\begin{array}{c} v_r \\ v_{\theta} \\ v_{\phi} \\  \end{array} \right]  = 
\left[\begin{array}{ccc}  \sin\bar{\theta}\cos\bar{\phi}  & \sin\bar{\theta}\sin\bar{\phi}  &  \cos\bar{\theta} \\
    \cos\bar{\theta}\cos\bar{\phi}  & \cos\bar{\theta}\sin\bar{\phi}  &  -\sin\bar{\theta}     \\
    -\sin\bar{\phi}  & \cos\bar{\phi}  &   0     \\
    \end{array} \right] \left[\begin{array}{c} v_{X} \\ v_{Y} \\ v_{Z} \\  \end{array} \right],  
    \label{eq:coordtrans_inverse}  
\end{equation}
where $\bar{\theta}=(\theta_A+\theta_B)/2$ and $\bar{\phi}=(\phi_A+\phi_B)/2$. Then, the perspective effects in the observed PMs and RVs can be corrected for by modifying the values of one star, e.g.\ star~A, as follows: 
\begin{equation}
\left. \begin{array}{ccl} \mu^\star_{\alpha,A} &   \rightarrow &  \mu^\star_{\alpha,B} + {v_\phi}/{(f\times \bar{d})}   \\
  \mu_{\delta,A} &   \rightarrow &  \mu_{\delta,B} - {v_\theta}/{(f\times \bar{d})}        \\
    {\rm RV}_A &  \rightarrow & {\rm RV}_B+v_r  \\   \end{array} \right\} ,  
    \label{eq:coordtrans_inverse}  
\end{equation}
where $\bar{d}=(d_A+d_B)/2$ (in practice we use the error-weighted mean which makes a negligible difference). This method returns the velocity of Proxima relative to $\alpha$-Cen AB in the $x^\prime y^\prime z^\prime$ frame of Figure\ref{fig:Euler2angles} as follows: $v_{x^\prime}=0.023\pm 0.025\,{\rm km}\,{\rm s}^{-1}$, $v_{y^\prime}=0.188\pm 0.027\,{\rm km}\,{\rm s}^{-1}$, and $v_{z^\prime}=0.196\pm 0.032\,{\rm km}\,{\rm s}^{-1}$. The magnitude of the 3D velocity is $v=0.273\pm 0.030\,{\rm km}\,{\rm s}^{-1}$ that agrees well with $v=0.273\pm 0.049\,{\rm km}\,{\rm s}^{-1}$ given by \cite{Kervella:2017}. However, we note that the error of $v$ quoted by \cite{Kervella:2017} corresponds to $\sqrt{\sigma_{v_{x^\prime}}^2+\sigma_{v_{y^\prime}}^2+\sigma_{v_{z^\prime}}^2}$ while the error estimated here corresponds to $\sqrt{\sigma_{v_{x^\prime}}^2{v_{x^\prime}^2}+\sigma_{v_{y^\prime}}^2{v_{y^\prime}}^2+\sigma_{v_{z^\prime}}^2{v_{z^\prime}}^2}/v$. This difference is irrelevant because only the velocity components with equivalent errors are used in both modeling procedures.

\end{document}